\documentclass[aps,prl,reprint,amsmath,amssymb,groupedaddress]{revtex4-2}
\usepackage{graphicx}
\usepackage[colorlinks,
linkcolor=blue,
anchorcolor=blue,
citecolor=blue,urlcolor=blue]{hyperref}
\begin{document}
	
	\title{Exact Resonances Are Not Sufficient for Phonon Energy Diffusion}
	
	\author{Wei Lin}
	\author{Yong Zhang}
	\author{Hong Zhao}
	\email{zhaoh@xmu.edu.cn}
	\affiliation{Department of Physics, Xiamen University, Xiamen 361005, Fujian, China}
	
	\date{\today}
	
	\begin{abstract}
		Multi-phonon resonance conditions underpin kinetic theories of phonon transport and lattice thermalization. We show that exact resonance matching, nonzero interaction coefficients, and network connectivity do not guarantee persistent energy diffusion. Symmetry-enforced balance relations drive exact-resonant collision currents to nonthermal zero-flux states, producing kinetic arrest from individual resonant sets to connected networks. Complete energy spreading is sustained by quasi-resonances. The thermodynamic and weak-nonlinearity limits do not commute: the leading kinetic behavior is recovered in the former, whereas at fixed finite size the thermalization time diverges through higher-order crossovers as the nonlinearity vanishes. Exact-resonance existence and connectivity are therefore kinematic, not sufficient dynamical, criteria for phonon energy diffusion.
	\end{abstract}
	
	\maketitle
	
	\section{I. Introduction}
	
	In weakly anharmonic lattices, vibrational energy is redistributed through interactions among normal modes, commonly described as phonon--phonon scattering within Peierls--Boltzmann kinetic theory~\cite{Peierls:1929,Dalitz:1997,Spohn:2006}. This framework provides a standard microscopic foundation for thermal-transport theory in crystalline solids, connecting anharmonic lattice dynamics to macroscopic heat conduction~\cite{Aoki:2006,Lukkarinen:2016}, and enabling quantitative calculations for bulk, low-dimensional, and nanoscale materials~\cite{Lindsay:2016,McGaughey:2019,Gu:2018}.  More broadly, resonance-based kinetic theories provide a fundamental route from microscopic mode coupling to irreversible transport and thermalization in weakly nonlinear many-body systems~\cite{Nazarenko:2011,Onorato:2023}. Recent mathematical advances have placed this framework on a first-principles footing by deriving the wave kinetic equation from nonlinear dispersive dynamics in joint large-system and weak-nonlinearity limits, with rigorous results covering the kinetic timescale, higher-order statistics, broad scaling regimes, and long-time evolution~\cite{LukkarinenSpohn:2011,Buckmaster:2021,Deng:2021,Deng:2023,Deng:2026Scaling,Deng:2026Propagation,Deng:2024}. The allowed scattering processes satisfy energy and crystal-momentum conservation in the standard multi-wave resonance form
	\begin{equation}
		\begin{aligned}
			\pm &k_1\pm k_2\pm\cdots\pm k_n
			=0\pmod{N},
			\\
			\pm&\omega_{k_1}\pm\omega_{k_2}\pm\cdots\pm\omega_{k_n}
			=0.
		\end{aligned}
		\label{eq:kw}
	\end{equation}
	The same sign pattern in both equations specifies a particular scattering process. These conditions, together with the phonon dispersion, determine which scattering processes are kinematically allowed~\cite{Ravichandran:2020,Ravichandran:2021}, while crystal symmetry imposes the further requirement that the corresponding interaction coefficient be nonzero~\cite{Gu:2018,Yang:2021}. Once resonance matching and nonzero coupling are established, such processes are conventionally retained as active scattering channels, and a connected network of these channels is commonly taken to provide a pathway to global spectral mixing and thermalization.
	
	Here we show that this textbook-level picture is incomplete: although resonance matching and nonzero coupling establish a channel as kinematically allowed, they do not guarantee a persistent collision current.   Even when an exact-resonant process is the leading one within a connected network, symmetry-enforced balance relations can drive its collision current to zero before equipartition, leaving the system in a nonthermal state. We term this kinetic arrest. This arrest persists across individual exact-resonant sets, coupled clusters, and connected exact-resonance networks. 
	
	More generally, exact resonance conditions remain valid kinematic constraints, but resonance matching, nonzero coupling, and network connectivity are not sufficient dynamical criteria for persistent energy diffusion in weakly nonlinear lattices. This distinction has fundamental implications for resonance-based kinetic theories and bears directly on longstanding problems in nonlinear lattice dynamics, including the Fermi--Pasta--Ulam--Tsingou (FPUT) problem~\cite{Fermi:1955,Campbell:2005,Gallavotti:2008}, a canonical setting for understanding how irreversible statistical behavior emerges from microscopic Hamiltonian dynamics. In finite-size periodic FPUT chains, lower-order exact resonances may fail to mix the full mode spectrum, whereas six-wave exact resonances can connect all normal modes into a resonant network~\cite{Onorato:2015,Lvov:2018}. This connectivity has been interpreted as a route to equipartition, motivating the expectation that finite-size chains thermalize even at arbitrarily weak nonlinearity and inspiring related exact-resonance scenarios for broader classes of weakly nonlinear lattices.
	
	Whether this connected network is dynamically sufficient must therefore be established rather than inferred from its topology. We show that it is not: the network generates a genuine kinetic transient, but its collision flow undergoes kinetic arrest before equipartition. Thermalization in FPUT and related finite-size lattices must therefore be understood in terms of the resonant channels that remain dynamically active. Quasi-resonances generated by nonlinear frequency broadening, rather than the discrete exact-resonance set alone, sustain the transport required for complete energy spreading and finite-time thermalization. The role of quasi-resonant interactions in finite and discrete wave systems has long been recognized~\cite{Nazarenko:2011,Connaughton:2001,Kartashova:2007,Lvov:2010,Pan:2017,Buckmaster:2021,Deng:2021,Deng:2023,Deng:2026Scaling}, and such interactions have been invoked to explain thermalization in both large lattices~\cite{Fu:2019,Fu:2019R,Wang:2020,Pistone:2019,Wang:2024} and multimode optical fibers~\cite{Garnier:2019,Baudin:2023,Chao:2025,Zanaglia:2026}.
	
	Building on these studies, we use the weighted connection strength introduced in Ref.~\cite{Lin:2025} to compare exact- and quasi-resonant channels on equal footing. At fixed broadening width, their strengths exhibit opposite size dependences: the discrete exact-resonant contribution decreases with system size, whereas the quasi-resonant contribution grows as the spectrum becomes denser. This growth is consistent with rigorous wave-kinetic limits, in which finite-volume quasi-resonant sums converge to a continuum collision operator supported on the exact-resonance manifold~\cite{Buckmaster:2021,Deng:2021,Deng:2023,Deng:2026Scaling}. These trends imply that the thermodynamic and weak-nonlinearity limits do not commute. Increasing $N$ at fixed nonzero nonlinearity strengthens the leading quasi-resonant network and recovers the leading kinetic behavior, whereas decreasing the nonlinearity at fixed $N$ narrows the frequency-broadening window and weakens the leading channel, driving crossovers to progressively higher-order quasi-resonant processes and causing the thermalization time to diverge. Exact-resonant transients emerge in crossover regimes where the leading quasi-resonant channel has weakened but higher-order processes have not yet become dominant.
	
	\section{II. Model and exact-resonant kinetics}
	
	Six-wave exact resonances are the lowest-order processes capable of connecting the full normal-mode spectrum in finite-size lattices~\cite{Onorato:2015,Lvov:2018}. To isolate the dynamical role of this network, we consider the energy-rescaled periodic FPUT-6 lattice
	\begin{equation}
		\tilde H=
		\sum_{j=0}^{N-1}
		\left[
		\frac{\tilde p_j^2}{2}
		+\frac{1}{2}(\tilde q_{j+1}-\tilde q_j)^2
		+\frac{g}{6}(\tilde q_{j+1}-\tilde q_j)^6
		\right],
		\qquad
		\\
		g=b\epsilon^2 ,
		\label{eq:H6_rescaled}
	\end{equation}
	where $\epsilon=H/N$ is the energy density, $g$ is the dimensionless nonlinear strength, and periodic boundary conditions are imposed. The pure sextic interaction makes six-wave scattering the leading nontrivial process, allowing a controlled test of the network's dynamical sufficiency without lower-order nonlinear scattering channels. Higher orders of the normal-form expansion generate effective ten-wave and higher-wave processes; see the Supplemental Material (SM)~\cite{SM}.
	
	The harmonic modes have frequencies $\omega_k=2|\sin(\pi k/N)|$, wave actions $D_k=\langle |a_k|^2\rangle$, and modal energies $E_k=\omega_kD_k$. Standard perturbation theory yields the leading six-wave kinetic equation, whose collision integral comprises the $1\leftrightarrow5$, $2\leftrightarrow4$, $3\leftrightarrow3$, $4\leftrightarrow2$, $5\leftrightarrow1$, and $6\leftrightarrow0$ partitions. In the weakly nonlinear regime considered here, the $1\leftrightarrow5$, $5\leftrightarrow1$, and $6\leftrightarrow0$ channels have no nontrivial exact solutions, while the $2\leftrightarrow4$ and $4\leftrightarrow2$ channels are either absent or restricted to sparse exceptional families that do not affect the kinetic arrest. 
	These exceptional channels are analyzed in the SM~\cite{SM}; for the lattice sizes considered below, they are absent, so the kinetic equation reduces to the $3\leftrightarrow3$ contribution:
	\begin{equation}
		\begin{aligned}
			\dot D_1
			=&
			-2400\pi g^2
			\sum_{k_2,\ldots,k_6\ne 0}
			\left|A_{123456}\right|^2
			\left(\prod_{i=1}^{6}D_i\right)
			\mathcal B\,
			\delta(\Delta\omega)
			\\
			=&
			\eta_1-\gamma_1D_1 .
		\end{aligned}
		\label{eq:compact_kinetic}
	\end{equation}
	Here $D_i\equiv D_{k_i}$ and $\omega_i\equiv\omega_{k_i}$, with
	\[
	|A_{123456}|
	= \frac{\sqrt{\prod_{i=1}^{6}\omega_i}}{64N^2}\,\delta_K ,
	\]
	where $\delta_K$ enforces
	$k_1+k_2+k_3=k_4+k_5+k_6\pmod N$. Moreover, $\Delta\omega=\omega_1+\omega_2+\omega_3-\omega_4-\omega_5-\omega_6$,
	and $\mathcal B = \frac{1}{D_4}+\frac{1}{D_5}+\frac{1}{D_6} -\frac{1}{D_1}-\frac{1}{D_2}-\frac{1}{D_3}$. The zero mode is excluded because it has $\omega_0=0$ and does not participate in the phonon kinetic evolution.
	When its collision integral is active, Eq.~\eqref{eq:compact_kinetic} gives the conventional six-wave kinetic timescale $T_{\mathrm{kin}}\sim g^{-2}$. A derivation of Eq.~\eqref{eq:compact_kinetic} is provided in the SM~\cite{SM}.
	
	We consider lattice sizes $N\not\equiv0\pmod3$, for which the relevant nontrivial $3\leftrightarrow3$ exact-resonance solutions fall into the symmetric and quasi-symmetric families~\cite{Onorato:2015,Bustamante:2019}
	\begin{align}
		(k_1,k_2,k_3)
		&\leftrightarrow
		(-k_1,-k_2,-k_3), &
		k_1+k_2+k_3=\ell N/2 ,
		\label{eq:sym33}
		\\
		(k_1,k_2,k_3)
		&\leftrightarrow
		(-k_1,-k_2,k_3), &
		k_1+k_2=\ell N/2,
		\label{eq:quasisym33}
	\end{align}
	where $\ell$ is an integer. These resonant families form a connected network linking all normal modes, which would suggest equipartition if connectivity were dynamically sufficient.
	
	However, both families obey the additional pairwise constraint
	\begin{equation}
		\frac{d}{dt}\left(D_k+D_{-k}\right)=0 .
		\label{eq:pair_invariant}
	\end{equation}
	The full derivation is presented in the SM~\cite{SM}. Because Eq.~\eqref{eq:pair_invariant} holds for every resonant set, it also holds for the full connected $3\leftrightarrow3$ exact-resonance network. Since $\omega_k=\omega_{-k}$, Eq.~\eqref{eq:pair_invariant} also conserves each pair energy, restricting exact-resonant dynamics to redistribution within counterpropagating pairs and allowing $\mathcal{B}$, and hence the collision currents, to vanish before equipartition.

	\begin{figure}[t]
		\centering
		\includegraphics[width=\columnwidth]{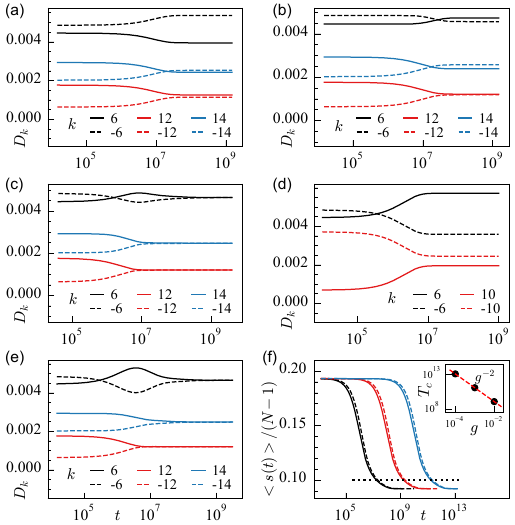}
		\caption{
			Direct numerical verification of exact-resonant kinetics in the periodic FPUT-6 chain.
			(a)--(e) Time evolution of the wave actions $D_k$ for $N=32$ and $g=10^{-2}$, retaining (a) a single symmetric sextet; (b) two coupled symmetric sextets; (c) all symmetric $3\leftrightarrow3$ exact resonances; (d) all quasi-symmetric $3\leftrightarrow3$ exact resonances; and (e) both families. 
			(f) Normalized indicator entropy $\langle s(t)\rangle/(N-1)$ for $g=10^{-2},10^{-3}$, and $10^{-4}$. The solid and dashed lines correspond, respectively, to calculations retaining both resonance families and the symmetric family alone.
			The inset shows the relaxation time $T_c$, defined by $\langle s(T_c)\rangle/(N-1)=0.1$; the dashed line indicates $T_c\propto g^{-2}$.
		} 
		
		\label{fig:exact_arrest}
	\end{figure}
	
	\section{III. Direct numerical test of exact-resonant kinetics.}
	
	We numerically integrate the exact-resonant kinetic equation while retaining different subsets of the $3\leftrightarrow3$ resonances. Each participating mode is assigned an independent initial energy uniformly sampled from $[0,0.006]$, giving $\langle E_k(0)\rangle=0.003$. Numerical details are provided in the SM~\cite{SM}. For the single symmetric sextet $(6,12,14)\leftrightarrow(-6,-12,-14)$ in Fig.~\ref{fig:exact_arrest}(a), the actions evolve toward a stationary state, but neither equipartition within the sextet nor equality between every pair $k$ and $-k$ is reached; some pairs even move farther from equality. Connecting a second symmetric sextet $(6,11,15)\leftrightarrow(-6,-11,-15)$ through the shared pair $(6,-6)$ does not alter this outcome [Fig.~\ref{fig:exact_arrest}(b)]. However, the complete symmetric network drives every counterpropagating pair toward equality before arresting [Fig.~\ref{fig:exact_arrest}(c)]. The quasi-symmetric network also arrests, generally without global pairwise equalization [Fig.~\ref{fig:exact_arrest}(d)], whereas combining the two families yields a pairwise-symmetric stationary state [Fig.~\ref{fig:exact_arrest}(e)].
	
	Defining the reciprocal pair imbalance as $\chi_k\equiv D_{-k}^{-1}-D_k^{-1}$, a single symmetric sextet imposes only $\chi_{k_1}+\chi_{k_2}+\chi_{k_3}=0$, allowing finite pair imbalances to cancel while conservation of each pair sum $D_k+D_{-k}$ restricts redistribution. Small connected clusters can therefore remain underconstrained. In the complete symmetric network, however, each pair belongs to multiple overlapping sextets and must satisfy several partner-dependent balance relations, which share the pairwise-symmetric zero-flux state $\chi_k=0$, equivalently $D_k=D_{-k}$.
	
	We measure the overall redistribution using the normalized indicator entropy $\langle s(t)\rangle/(N-1)$\cite{Onorato:2015,Lin:2025}, where $s(t)=\sum_{k\neq0}f_k\ln f_k$, $f_k=E_k/\bar E$, and $\bar E=(N-1)^{-1}\sum_{k\neq0}E_k$. Equipartition corresponds to $s=0$.  The initial uniform sampling yields $\langle s(0)\rangle/(N-1)\approx0.191$; pairwise equalization would lead to a plateau at $\langle s\rangle/(N-1)\simeq0.091$~\cite{SM}, in quantitative agreement with Fig.~\ref{fig:exact_arrest}(f). 
	This nonzero value reflects that only each counterpropagating pair is equalized, while different pairs retain distinct energies. The collapse as a function of $g^2t$ in Fig.~\ref{fig:exact_arrest}(f), together with the inset scaling $T_c\propto g^{-2}$, demonstrates a genuine exact-resonant $g^{-2}$ kinetic transient that arrests before equipartition. Its visibility in the full dynamics depends on competition with the quasi-resonant channels considered next.
	
	\section{IV. Size and nonlinearity dependence of resonant channels.}
\begin{figure}[t]
		\centering
		\includegraphics[width=\columnwidth]{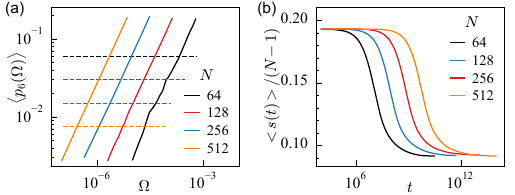}
		\caption{Size dependence of six-wave resonant channels.
			(a) Mode-averaged weighted connection strength $\langle p_6(\Omega)\rangle$ as a function of the broadening width $\Omega$ for $N=64,128,256$, and $512$. Solid curves show the quasi-resonant contribution from $0<\lvert\Delta\omega\rvert<\Omega$, whereas dashed horizontal lines show the exact-resonant contribution at $\Delta\omega=0$. 	(b) Normalized indicator entropy $\langle s(t)\rangle/(N-1)$ obtained from the exact-resonant kinetic equation including all allowed $3\leftrightarrow3$ processes at $g=10^{-2}$ for the same system sizes.}
		\label{fig:network_connectivity}
	\end{figure}

	To determine when the arrested exact-resonant transient can emerge in the full dynamics, we compare the exact- and quasi-resonant six-wave contributions. Nonlinear frequency broadening permits a finite frequency mismatch,
	\begin{equation}
		\lvert\Delta\omega\rvert=\lvert \omega_{k_1} + \omega_{k_2} + \omega_{k_3} - \omega_{k_4} - \omega_{k_5} - \omega_{k_6} \rvert < \Omega
		\label{eq:quasi_resonance}
	\end{equation}
	where $\Omega$ is the nonlinear broadening width. For the present model, $\Omega\sim g^2$~\cite{SM}. Following Ref.~\cite{Lin:2025}, we quantify the coupling of mode $k_1$ to the allowed six-wave processes by the weighted connection strength
	\begin{equation}
		p_6(k_1;\Omega)=
		\sum_{\mathcal Q_\Omega(k_1)}
		\left|A_{k_1k_2k_3k_4k_5k_6}\right|,
		\label{eq:p6}
	\end{equation}
	where $\mathcal Q_\Omega(k_1)$ contains the six-wave tuples satisfying the crystal-momentum condition and Eq.~\eqref{eq:quasi_resonance}. We evaluate Eq.~\eqref{eq:p6} separately over $\Delta\omega=0$ for exact resonances and $0<\lvert\Delta\omega\rvert<\Omega$ for quasi-resonances. Figure~\ref{fig:network_connectivity}(a) compares their mode averages, $\langle p_6(\Omega)\rangle=(N-1)^{-1}\sum_{k_1\neq0}p_6(k_1;\Omega)$.
	
	The exact- and quasi-resonant contributions exhibit opposite size dependences. At fixed $\Omega$, the quasi-resonant strength grows with $N$ due to a denser spectrum,  while the exact-resonant strength declines [Fig.~\ref{fig:network_connectivity}(a)]. The latter is a structural property independent of $g$, although the corresponding kinetic rate retains the overall factor $g^2$. Consistently, Fig.~\ref{fig:network_connectivity}(b) shows that exact-resonant relaxation becomes progressively slower with increasing $N$.
	
	These trends yield two predictions. At fixed nonzero $g$, the leading six-wave quasi-resonances control the thermodynamic limit, consistent with rigorous large-box derivations of wave kinetic theory~\cite{Buckmaster:2021,Deng:2021,Deng:2023,Deng:2026Scaling}; the opposite size dependences of the two connection strengths provide a direct network-level interpretation of this limit. Once the leading quasi-resonant contribution becomes sufficiently weak, relaxation can cross over to either the exact six-wave transient or a higher-order quasi-resonant process, depending on their relative strengths. 
	
	\section{V. Full-dynamics test of the resonant-channel hierarchy.}
	
\begin{figure}[tb]
		\centering
		\includegraphics[width=\columnwidth]{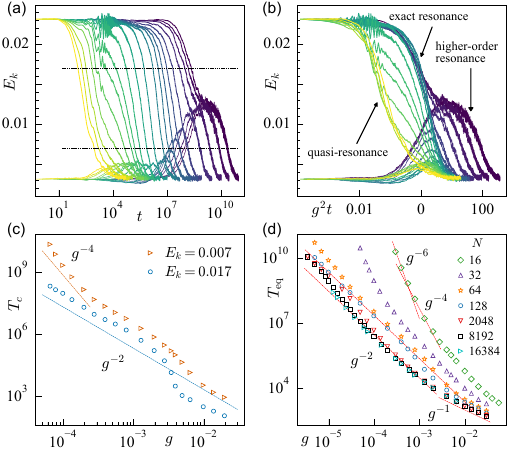}
		\caption{
			Full Hamiltonian dynamics of the periodic FPUT-6 chain.
			(a) Evolution of the $k=\pm1$ mode energies for N=32, with $g$ decreasing from $0.02$ to $6\times10^{-5}$. The horizontal lines mark the thresholds $E_{k=1}=0.017$ and $E_{k=1}=0.007$.
			(b) The same data plotted against the rescaled time $g^2t$.
			(c) Relaxation time $T_c$ as a function of $g$ for the two thresholds in (a).
			(d) Equipartition time $T_{\rm eq}$ for different system sizes. Dashed lines indicate reference power laws. }
		\label{fig:Ek-t-6p}
	\end{figure}

	We now test these predictions using the full Hamiltonian dynamics. Details of the Hamiltonian simulations are provided in the SM~\cite{SM}, while the thermalization time is computed following Ref.~\cite{Lin:2025}. Figure~\ref{fig:Ek-t-6p}(a) shows the initially excited $k=1$ mode and its counterpropagating partner for $N=32$ as $g$ decreases from $0.02$ to $6\times10^{-5}$. Two distinct energy-transfer patterns emerge. At relatively large $g$, the energy of the $k=1$ mode decreases rapidly, whereas the $k=-1$ mode receives only a small fraction of the lost energy; the remainder is redistributed to modes outside the pair through the connected six-wave quasi-resonant network. At smaller $g$, the dynamics crosses over to a predominantly pairwise exchange between $k=1$ and $k=-1$, characteristic of the exact six-wave kinetics. After the pair approaches equal energies, higher-order quasi-resonant processes in the full dynamics gradually transfer the pair's mean energy to the remaining modes. The $g^2t$ representation in Fig.~\ref{fig:Ek-t-6p}(b) clearly distinguishes these regimes.
	
	Figure~\ref{fig:Ek-t-6p}(c) locates these two six-wave regimes using the threshold time $T_c$. For $E_c=0.017$, the two $T_c\propto g^{-2}$ branches correspond to quasi-resonant redistribution (large $g$) and exact-resonant pairwise equalization (small$g$), showing identical scaling from different mechanisms. At $E_c=0.007$, the  small-$g$ branch departs from $g^{-2}$ since the arrested exact flow cannot reach this level; higher-order quasi-resonances are required.
	
	The equipartition times in Fig.~\ref{fig:Ek-t-6p}(d), defined by $\langle s(T_{\rm eq})\rangle/(N-1)=0.1$, organize this hierarchy across system size. At fixed nonzero $g$, increasing $N$ extends the weakly nonlinear $T_{\rm eq}\propto g^{-2}$ regime toward smaller  $g$, demonstrating that the denser six-wave quasi-resonant network increasingly controls relaxation toward the thermodynamic limit. At strong nonlinearity, the approximate $g^{-1}$ behavior lies outside weak-wave kinetics and is consistent with rapid chaotic mixing caused by nonlinear resonance overlap~\cite{Chirikov:1979,Lvov:2018}.
	
	For large finite systems, $N=2048$ and $8192$, the six-wave quasi-resonant $g^{-2}$ range broadens with $N$. As $g$ decreases, its first breakdown indicates a crossover to higher-order quasi-resonant relaxation, while the subsequent reentrant $g^{-2}$ segment signals that these higher-order channels have weakened below the exact six-wave contribution, allowing the exact-resonant transient to control the chosen threshold. Since this flow arrests, a stricter thermalization criterion would remove the reentrant segment. 
	For $N=64$ and $128$, decreasing $g$ drives successive crossovers from quasi-resonant six-wave relaxation to the exact six-wave transient and then to higher-order quasi-resonant relaxation. Because the first two mechanisms share the $g^{-2}$ timescale, their crossover appears as an apparently continuous, extended $g^{-2}$ regime, whose eventual breakdown marks the onset of the higher-order processes required for complete thermalization. For $N=16$ and $32$, the successive $g^{-2}$, $g^{-4}$, and $g^{-6}$ regimes are associated with effective six-, ten-, and fourteen-wave quasi-resonant processes, respectively, showing that small finite systems do not retain a persistent $g^{-2}$ thermalization law. The absence of a separately resolved exact six-wave transient is consistent with comparable strengths of the exact six-wave and higher-order quasi-resonant contributions in the crossover region.
	
	\section{VI. Generality of symmetry-constrained kinetic behavior}
	
	To assess the generality of symmetry-constrained kinetic behavior, we consider three additional settings: the discrete nonlinear Schr\"odinger (DNLS) model, the periodic FPUT-$\beta$ chain, and the fixed-boundary FPUT-6 chain. The DNLS model demonstrates that even a connected four-wave exact-resonance network can undergo kinetic arrest. The FPUT-$\beta$ chain shows that the same mechanism operates within mutually disconnected four-wave resonance quartets, each of which arrests independently. The fixed-boundary FPUT-6 chain further shows that the corresponding pairwise-symmetry structure can organize an early-time transient in the full Hamiltonian dynamics even when the normal-mode spectrum admits no nontrivial exact six-wave resonances. The theoretical analysis and numerical verification are presented below, while details of the numerical methodology are provided in the SM~\cite{SM}.
	
	\subsection{A. DNLS model.}
	
	We first consider the one-dimensional DNLS model. Schr\"odinger-type wave systems have served as a primary setting for rigorous derivations of wave kinetic theory. Early kinetic-limit results were established for weakly nonlinear lattice Schr\"odinger dynamics~\cite{LukkarinenSpohn:2011}, while subsequent works on the cubic NLS in large periodic domains rigorously established the onset of wave-kinetic behavior~\cite{Buckmaster:2021,Deng:2021}, its validity on the kinetic timescale~\cite{Deng:2023}, higher-order statistics and propagation of chaos~\cite{Deng:2026Propagation}, broad large-box and weak-coupling scaling regimes~\cite{Deng:2026Scaling}, and long-time validity over the lifespan of the corresponding kinetic solution~\cite{Deng:2024}. These results concern joint large-box and weak-coupling limits. In the scaling regimes that yield the wave kinetic equation, the vanishing mode spacing makes finite-time near-resonant interactions increasingly dense, and their cumulative contribution converges to a continuum four-wave collision operator. The large-box analysis also delineates scaling regimes in which discrete exact resonances dominate instead of the usual wave-kinetic limit~\cite{Deng:2026Scaling}. Here we address the complementary finite-system question: whether a connected network of discrete exact four-wave resonances is by itself sufficient to sustain kinetic redistribution.
	
	The DNLS model is governed by
	\begin{equation}\label{eq:dyn}
		i \frac{d\psi_l}{dt} = - (\psi_{l+1} + \psi_{l-1}) + V_l \psi_l + \beta |\psi_l|^2 \psi_l.
	\end{equation}
	Here, $\psi_l$ is the complex field amplitude at site $l=1,\ldots,N$. The coefficient $\beta$ characterizes the Kerr nonlinearity. We set the on-site potential to $V_l=0$ and impose fixed boundary conditions, $\psi_0=\psi_{N+1}=0$. This model provides a standard description of coupled nonlinear waveguide and multicore-fiber systems~\cite{Chekhovskoy:2016}. The lowest-order nontrivial scattering process is four-wave scattering, governed by the kinetic equation~\cite{Nazarenko:2019}
	\begin{equation}
		\label{eq:ke-nls}
		\dot{D}_{1}
		=-\frac{4\pi\beta^2}{(N+1)^2}
		\sum_{2,3,4}
		\left(\prod_{i=1}^{4}D_i\right)
		\mathcal B_{d}\delta_{Kd}\delta(\Delta\omega_d),
	\end{equation}
	where $\mathcal B_{d}=D_3^{-1}+D_4^{-1}-D_1^{-1}-D_2^{-1}$, $\Delta\omega_d=\omega_1+\omega_2-\omega_3-\omega_4$, and $\delta_{Kd}$ enforces $k_1+k_2=k_3+k_4\pmod{N+1}$.
	
	The linear normal modes are standing waves,
	\begin{align}
		\phi_k(l)
		&=\sqrt{\frac{2}{N+1}}
		\sin\left(\frac{\pi k l}{N+1}\right), \\
		\omega_k
		&=-2\cos\left(\frac{\pi k}{N+1}\right),
		\label{eq:DNLS_modes}
	\end{align}
	where \(k=1,\ldots,N\). For each mode $k$, define its complementary partner $\bar{k}=N+1-k$. Writing $q_k=\pi k/(N+1)$, one has $q_{\bar{k}}=\pi-q_k$ and hence
	\begin{equation}
		\omega_{\bar{k}}=-\omega_k,
		\qquad
		\phi_{\bar{k}}(l)=(-1)^{l+1}\phi_k(l).
		\label{eq:DNLS_staggered_pair}
	\end{equation}
	Thus, $k$ and $\bar{k}$ have the same standing-wave intensity profile but differ by a staggered phase between neighboring sites. They are staggered spectral partners associated with the bipartite symmetry of the linear DNLS chain, rather than the counterpropagating, equal-frequency modes $k$ and $-k$ in the periodic FPUT chains.
	
	This spectral symmetry generates a family of nontrivial exact four-wave resonances
	\begin{equation}
		k_1+k_2=k_3+k_4=N+1,
		\qquad
		\omega_{k_1}+\omega_{k_2}
		=\omega_{k_3}+\omega_{k_4}=0,
		\label{eq:DNLS_resonance_family}
	\end{equation}
	in which $(k_1,k_2)$ and $(k_3,k_4)$ are two staggered pairs. These quartets form a connected resonance network. Trivial permutations are excluded by requiring $k_1\ne k_3$ and $k_1\ne k_4$.
	
	After summing all quartets in the connected family, the two members of every staggered pair have equal net kinetic rates. For two representative pairs satisfying $k_1+k_2=k_3+k_4=N+1$,
	\begin{equation}
		\dot D_1=\dot D_2=J_{s0},
		\qquad
		\dot D_3=\dot D_4=J_{s1},
		\label{eq:pair_invariant1}
	\end{equation}
	so that $D_k-D_{\bar{k}}$ is conserved for every staggered pair. This differs from the periodic FPUT constraint in Eq.~\eqref{eq:pair_invariant}, which conserves the sum of the actions of two counterpropagating modes. Because each staggered pair participates in multiple quartets of the connected network, distinct staggered pairs generally evolve at different rates ($J_{s0}\neq J_{s1}$). Nevertheless, the same qualitative consequence follows: the constrained evolution stops once the connected resonance family reaches the zero-flux condition $\mathcal B_d=0$ for all active quartets.
	
\begin{figure}[tb]		
		\centering
		\includegraphics[width=\columnwidth]{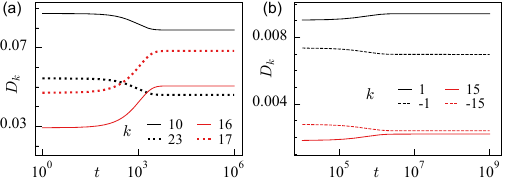}
		\caption{Symmetry-constrained kinetic arrest in the DNLS model and the periodic FPUT-$\beta$ chain. Exact four-wave kinetic evolution of $D_k$ for (a) the DNLS model and (b) the periodic FPUT-$\beta$ chain, both with $N=32$.}\label{fig:EM_beta_DNLS}
	\end{figure}

	Figure~\ref{fig:EM_beta_DNLS}(a) shows the numerical solution of Eq.~\eqref{eq:ke-nls}, confirming this analysis. Before the zero-flux state is reached, the staggered partners $k_1=10$ and $k_2=23$, which satisfy $k_1+k_2=N+1$, evolve at the same rate $J_{s0}$, while $k_3=16$ and $k_4=17$ evolve at the same rate $J_{s1}$. The two pairs evolve in opposite directions with unequal slope magnitudes, confirming that $J_{s0}\neq J_{s1}$. For $t\gtrsim10^4$, the wave actions of all four modes become time independent, indicating that the exact-resonant flow has reached the zero-flux state. Thus, the DNLS model exhibits a genuine exact-resonant transient followed by symmetry-constrained kinetic arrest. This result extends the mechanism beyond FPUT lattices while showing that its underlying modal symmetry can be physically distinct: staggered spectral pairing in the DNLS model replaces counterpropagating-mode pairing in the periodic FPUT chains.
	
	\subsection{B. Periodic FPUT-$\beta$ chain.}
	
	We next consider the periodic FPUT-$\beta$ chain,
	\begin{equation}
		\tilde H_{\beta}=\sum_j\left[
		\frac{\tilde p_j^2}{2}
		+\frac{1}{2}(\tilde q_j-\tilde q_{j+1})^2
		+\frac{g}{4}(\tilde q_j-\tilde q_{j+1})^4
		\right], 
		\label{eq:EM_FPUT_beta}
	\end{equation}
	where \(g=\beta\epsilon\). Its leading nontrivial process is four-wave scattering, governed by
	\begin{equation}
		\begin{aligned}
			\dot D_1=&-\frac{9\pi g^2}{32N^2}
			\sum_{k_2,k_3,k_4}
			\left(\prod_{i=1}^{4}\omega_iD_i\right)
			\mathcal B_{\beta}\delta_{K\beta}\delta(\Delta\omega_\beta).
		\end{aligned}
		\label{eq:EM_beta_kinetic}
	\end{equation}
	Here, $\mathcal B_{\beta}=D_3^{-1}+D_4^{-1}-D_1^{-1}-D_2^{-1}$, $\Delta\omega_\beta=\omega_1+\omega_2-\omega_3-\omega_4$, and $\delta_{K\beta}$ enforces $k_1+k_2=k_3+k_4\pmod N$. The nontrivial exact solutions are pairing quartets
	\begin{equation}
		(k_1,k_2)\leftrightarrow(-k_1,-k_2),
		\qquad k_1+k_2=\ell N/2.
		\label{eq:EM_beta_quartet}
	\end{equation}
	As established by the resonance classification in Ref.~\cite{Onorato:2015}, these quartets are mutually disjoint and therefore do not form a connected network spanning the full spectrum. Nevertheless, each quartet supports a symmetry-constrained exact-resonant current until its own zero-flux condition is reached, at which point it arrests independently.
	
	For a single quartet, Eq.~\eqref{eq:EM_beta_kinetic} gives
	\begin{equation}
		\dot D_{1}=\dot D_{2}=J_4,
		\qquad
		\dot D_{-1}=\dot D_{-2}=-J_4,
		\label{eq:EM_beta_current}
	\end{equation}
	so that the pairwise constraint in Eq.~\eqref{eq:pair_invariant} also holds in the FPUT-$\beta$ model. The current vanishes when
	\begin{equation}
		\frac{1}{D_{-1}}+\frac{1}{D_{-2}}
		-\frac{1}{D_{1}}-\frac{1}{D_{2}}=0,
		\label{eq:EM_beta_zero_flux}
	\end{equation}
	which requires neither equipartition nor pairwise equality. Figure~\ref{fig:EM_beta_DNLS}(b) confirms that the evolution follows the constrained current in Eq.~\eqref{eq:EM_beta_current} and arrests upon reaching the zero-flux state in Eq.~\eqref{eq:EM_beta_zero_flux}. Thus, symmetry-induced kinetic arrest also occurs in a four-wave FPUT lattice despite the absence of a globally connected exact-resonance network.
	
	\subsection{C. Fixed-boundary FPUT-6 chain.}
	
	We finally examine whether the symmetry-constrained transient persists under fixed boundary conditions. The normal modes of the fixed-boundary FPUT-6 chain are standing waves, and their dispersion admits no nontrivial exact six-wave resonances. Nevertheless, the field can be decomposed into counterpropagating traveling-wave components with wavenumbers $k$ and $-k$. Although these components are not independent normal modes, the pairwise symmetry inherited from the periodic-chain resonant structure can still organize their early-time dynamics, producing a transient tendency toward counterpropagating-mode equalization before broader redistribution sets in.

	\begin{figure}[htb]
		\centering
		\includegraphics[width=\columnwidth]{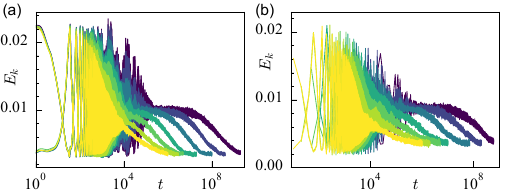}
		\caption{
			Symmetry-constrained transient in the fixed-boundary FPUT-6 lattice. (a,b) Time evolution of the $k=\pm1$ traveling-wave component energies for $N=32$ and $64$, respectively. The colored curves, from left to right, correspond to decreasing nonlinearity strengths from $g=3\times10^{-3}$ to $10^{-4}$. As $g$ decreases, the early-time dynamics increasingly approaches pairwise balance, followed by a slower redistribution beyond the pairwise transient.}\label{fig:EM_FPUT6fixed}
	\end{figure}
	
	To test this behavior, we initialize the projected $k$ and $-k$ components with unequal amplitudes and monitor their evolution under the full Hamiltonian dynamics. Figures~\ref{fig:EM_FPUT6fixed}(a) and \ref{fig:EM_FPUT6fixed}(b) show the results for $N=32$ and $64$, respectively. Boundary reflections produce pronounced oscillations in the component energies, but the early-time dynamics still approaches counterpropagating-mode balance. For the sizes considered, the equalization time increases with $N$ and is shorter than in the periodic chain, after which slower redistribution transfers energy beyond the pair. Thus, fixed boundaries modify the temporal profile but do not eliminate the symmetry-constrained transient.
	
	\section{VII. Discussion and conclusions}
	
	In summary, the existence and connectivity of exact multi-phonon resonances are only kinematic criteria for phonon energy diffusion: symmetry-enforced balance relations can arrest the associated collision currents at nonthermal zero-flux states, so complete energy spreading requires dynamically active quasi-resonant processes. This revised dynamical criterion reframes the microscopic picture of lattice energy diffusion, explains how the continuum wave-kinetic limit emerges from increasingly dense quasi-resonant interactions despite the arrest of discrete exact-resonant flows at fixed system size. Broadly speaking, this kinetic arrest parallels the nonthermal energy localization known from discrete breathers or KAM tori ~\cite{Flach:2008,Flach:2005}, reinforcing that kinematic resonance criteria alone are insufficient to ensure ergodicity. Crucially, the thermodynamic limit and the weak-nonlinearity limit do not commute, implying that the standard wave-kinetic description must be interpreted with caution when applied to finite systems.   
	
	\bibliography{ref}
	
\end{document}